\documentstyle[aps,prl,epsfig,twocolumn]{revtex}

\def\be{\begin{equation}}
\def\ee{\end{equation}}
\def\bea{\begin{eqnarray}}
\def\eea{\end{eqnarray}}
\def\bma{\begin{mathletters}}
\def\ema{\end{mathletters}}

\begin{document}

\draft

\title{Optimal purification of single qubits}

\author{J. I. Cirac,$^1$ A. K. Ekert,$^2$ and C. Macchiavello$^3$}

\address{$^{1}$Institut f\"{u}r Theoretische Physik, Universit\"{a}t Innsbruck,
Technikerstr. 25, A-6020 Innsbruck, Austria.\\
$^{2}$ Department of Physics, Clarendon Laboratory, University of Oxford,
Oxford OX1 3PU, U.K.\\
$^{3}$ Dipartimento di Fisica ``A. Volta'' and INFM--Unit\`{a} di Pavia, Via
Bassi 6, 27100 Pavia, Italy}

\date{\today}

\maketitle

\begin{abstract}
We introduce a new decomposition of the multiqubit states of the form
$\rho^{\otimes N}$ and employ it to construct the optimal single qubit
purification procedure. The same decomposition allows us to study
optimal quantum cloning and state estimation of mixed states.
\end{abstract}


\bigskip 

Quantum data processing \cite{comp}, such as quantum cryptography,
teleportation or computation, rely on an ample supply of qubits in
nearly pure quantum states. However, maintaining unknown and uncorrupted
entanglement or superpositions in the presence of the coupling to the
environment is usually a difficult task and a great deal of recent
research has been focused on issues such as entanglement purification
\cite{purif}. In particular, several entanglement purification protocols
have been devised but so far none of them has been proved to be optimal;
that is, it is not clear what is the maximal fraction of purified
entangled particles for a given initial state and a prescribed degree of
purification. This paper addresses a less complicated question of
purifying unknown quantum superpositions of qubits; however, it provides
a definite and constructive answer regarding the optimality of the
single-qubit purification procedures. The solution of the problem
involves a particular decomposition of density operators of multiqubit
systems. This decomposition seems to be a useful tool on its own as it
also gives some insight into the related problems of the optimal cloning
\cite{Gi97} and the optimal state estimation \cite{Ma95} of mixed
states.

Suppose we are given $N$ identical qubits, each in an unknown state
described by a density operator 
\begin{equation}
\rho =\frac{1}{2}\left( {\bf 1}+\lambda \vec{n}\cdot \vec{\sigma}\right)
= c_{1}|1_{\vec{n}}\rangle\langle 1_{\vec{n}}| +
  c_{0}|0_{\vec{n}}\rangle\langle 0_{\vec{n}}|,
\label{rhop}
\end{equation}
where $\lambda \vec{n}$ is the Bloch vector, $|\vec{n}|=1$,
$c_{1}-c_{0}=\lambda >0$, $c_{1}+c_{0}=1$ and $|0_{\vec{n}}\rangle,
|1_{\vec{n}}\rangle$ are the two eigenstates of $\rho$. The goal is to
obtain the maximum number of qubits in a state ``as close as possible''
to the unknown state $|1_{\vec{n}}\rangle$, i.e. to {\em purify the
state} $\rho$. Before specifying more precisely the problem, let us
re--write $\rho$ as a convex sum
\begin{equation}
\rho =\lambda \ \frac{1}{2}\left( {\bf 1}+\vec{n}\cdot \vec{\sigma}\right)
+(1-\lambda )\ \frac{1}{2}{\bf 1}
\end{equation}
where the first term on the r.h.s. represents the purified state
$\frac{1}{2}\left({\bf 1}+\vec{n}\cdot \vec{\sigma}\right) =
|1_{\vec{n}}\rangle\langle 1_{\vec{n}}|$ and the second one the
maximally mixed state $\frac{1}{2}{\bf 1}$. This decomposition implies
that given a set of $N$ identically and independently prepared qubits
one can distill from the set at most $\lambda N$ of perfectly purified
qubits. In the following we show that this limit is achievable when
$N\to \infty$ and we describe the optimal purification procedure for any
finite $N$.

In general, a purification procedure ${\cal P}$ consists of a set of
operations and measurements on the $N$ qubits and perhaps on additional
ancillas. Depending on the outcomes of the measurements some of the
qubits will be discarded and we will end up with $M\le N$ qubits whose
reduced density operators are closer to the ideal state
$|1_{\vec{n}}\rangle$. Thus ${\cal P}$ is described by a set of linear
completely positive maps ${\cal P}_{M}$ such that ${\cal
P}_{M}(\rho^{\otimes N})=\rho _{M}$ where $\rho _{M}$ is an unnormalized
density operator describing $M$ qubits. The probability of obtaining $M$
qubits is $P_{M}={\rm tr}(\rho _{M})$ and their corresponding fidelities
$F_{M}^{k}=\langle 1_{\vec{n}}|\rho_{k}|1_{\vec{n}}\rangle$, where
$\rho_{k}$ is the normalized reduced density operator of qubits
$k=1,2,\ldots,M$. These probabilities and fidelities depend both on the
direction $\vec n$ and the length $\lambda$ of the Bloch vector of the original 
qubits, and thus we write them as $P_{M}(\lambda,\vec{n})$ and
$F^k_{M}(\lambda,\vec{n})$. Given that we are not interested in the
fidelity of a particular qubit, nor in a particular direction of the Bloch
vector but rather in their average values, we define 
\begin{mathletters}
\label{PMFM}
\begin{eqnarray}
P_{M}(\lambda) &=&\int d\vec{n} \;P_{M}(\lambda ,\vec{n}), \\
\label{FM}
F_{M}(\lambda) &=&\frac{1}{P_{M}(\lambda )}\int d\vec{n} 
  \; P_{M}(\lambda ,\vec{n})
  \frac{1}{M} \sum_{k=1}^{M}F_{M}^k(\lambda,\vec{n}),
\end{eqnarray}
\end{mathletters}
where the integral refers to the complete solid angle.
Thus, any given ${\cal P}$ is characterized by these two quantities
namely, the probability of obtaining $M$ valid qubits and their
corresponding fidelity. This fact implies that there is no unique way of
defining optimality. For example, a procedure might give with small
probability a large number of qubits of high fidelity, whereas another
one might give with a large probability a smaller number of qubits of
higher fidelity. Nevertheless, it turns out that there is a procedure
${\cal P}$ which has the following unique properties: (1) given any
other purification procedure ${\cal P}'$ characterized by $P_M'$ and
$F_M'$, it is always possible to obtain these values by first applying
${\cal P}$ and then some prescribed operations and measurements; (2)
after applying ${\cal P}$, it is impossible to increase the fidelity of
the outcoming qubits even at the cost of decreasing the number of
qubits. The first property is a consequence of the fact that ${\cal P}$
is, in a certain sense, reversible. It implies that after applying it,
we can still obtain the optimal results whatever we mean by optimal. The
second property implies that after applying ${\cal P}$, it is impossible
to purify further. As a consequence, our procedure is optimal for any
reasonable definition of optimality. On the other hand, we introduce now
an important concept that will be used later on to prove optimality of
our protocol. We say that a procedure ${\cal P}^s$ is {\em symmetric}
if: (i) the reduced density operators of the $M$ qubits are identical;
(ii) the maps ${\cal P}_{M}^{s}$ are covariant \cite{We98}, that is, 
\begin{equation}
{\cal P}_{M}^{s}\left[ (U\rho U^{\dagger })^{\otimes N}\right] =U^{\otimes M}%
{\cal P}_{M}^{s}\left( \rho ^{\otimes N}\right) (U^{\dagger })^{\otimes M}
\label{cov}
\end{equation}
for all $U\in SU(2)$. These conditions imply that the probability of
obtaining $M$ qubits is independent of $\vec{n}$ and that the fidelities of
all the qubits are the same, which simplifies the calculations. Moreover,
they are the only ones that have to be analyzed since for any arbitrary
purification protocol ${\cal P}$ one can construct a symmetric one which
gives exactly the same values of $P_{M}(\lambda)$ and $F_{M}(\lambda )$. In
order to do that, we proceed as follows: (1) we apply the same random
unitary $U$ operator to the $N$ qubits; (2) we apply the procedure ${\cal P}$; 
(3) we apply the operator $U^{\dagger }$ to the $M$ resulting qubits; (4)
we permute the $M$ qubits randomly. 

We shall start with a set of $N=2J$ qubits (the $2J+1$ case can be
solved in an analogous way) and decompose their state $\rho^{\otimes N}$
into the sum of density operators $\rho_{j,\alpha}$ which are supported
on orthogonal subspaces $S_{j,\alpha_j}={\rm span}\{|j,m,\alpha_j
\rangle;$ $m=-j,-j+1,\ldots,j\}$, where $j=0,1,\ldots,J$,
$\alpha_j=1,2,\ldots,d_j$, and 
\begin{equation}
d_j = \left( \begin{array}{c} 2J \\ J-j \end{array}\right)
  - \left( \begin{array}{c} 2J \\ J-j-1 \end{array}\right) \quad (j\ne J),
\end{equation}
$d_J=1$. States $|j,m,\alpha\rangle$ are constructed from an arbitrary
basis $\{|0\rangle,|1\rangle\}$ of a single qubit in such a way that:
(i) for a fixed $j$ and $\alpha$, vectors $|j,m,\alpha \rangle$ form a
basis for the $(2j+1)$--dimensional irreducible representation of
$SU(2)$; (ii) for a fixed $j$ and $m$, vectors $|j,m,\alpha \rangle $
form a basis for the $d_{j}$--dimensional representation of $S_{N}$
(symmetric group) corresponding to the Young diagram $(2j,2J-2j)$. In
particular, we choose 
\begin{equation}
|j,m,1\rangle =|j,m\rangle \otimes |\Psi _{-}\rangle ^{\otimes J-j},
\end{equation}
where $|j,m\rangle$ is the symmetric state of $2j$ qubits with $j-m$ qubits
in the state $|0\rangle $ and $j+m$ qubits in the state $|1\rangle$, 
and $|\Psi _{-}\rangle$ is the singlet state of two qubits. The rest of the
states are constructed by using permutation operators, $\Pi_{i}\in S_{N}$, 
\begin{equation}
|j,m,\alpha \rangle =\sum_{i}h_{i}\Pi _{i}|j,m,1\rangle  
 \label{permutation}
\end{equation}
and the coefficients $h_{i}$ are chosen so that the corresponding states
are orthonormal. Although we have defined the basis $|j,m,\alpha\rangle$
in terms of the states $|i\rangle$ ($i=0,1$), we could have used any
other basis $|i_{\vec n}\rangle= U_{\vec n}|i\rangle$ where $U_{\vec
n}\in SU(2)$ to define $|j,m,\alpha\rangle_{\vec n}$. One can easily
check that
\begin{equation}
\label{irrrep}
|j,m,\alpha\rangle_{\vec n} = U_{\vec n}^{\otimes N}|j,m,\alpha\rangle =
\sum_{m'} D^j_{m',m}(\vec n) |j,m',\alpha\rangle,
\end{equation}
where $D^j_{m',m}(\vec n)$ are the corresponding matrix elements. Using
(\ref{irrrep}) one immediately sees that the subspaces $S_{j,\alpha}$
are the same if we take any basis $|j,m,\alpha\rangle_{\vec n}$ to
construct them. We also define a set of unitary operators $U_{j,\alpha}$
which transform $S_{j,\alpha}$ into $S_{j,1}$ while leaving the other
subspaces untouched. We take $U_{j,\alpha}|j,m,\alpha\rangle=
|j,m,1\rangle$ and $U_{j,\alpha}|j,m,1\rangle=|j,m,\alpha\rangle$.
Again, using (\ref{irrrep}) one can easily check that these definitions
are valid for all $\vec n$, i.e. $U_{j,\alpha}|j,m,1\rangle_{\vec n}=
|j,m,\alpha\rangle_{\vec n}$. 

The decomposition itself reads
\begin{equation}
\label{rhoN}
\rho ^{\otimes N}=\sum_{j=0}^{J}p_{j}\frac{1}{d_{j}}
  \sum_{\alpha=1}^{d_{j}}\rho _{j,\alpha },  
\end{equation}
where 
\begin{mathletters}
\begin{eqnarray}
\label{Pj}
p_{j} &=& d_{j}[c_{0}c_{1}]^{J-j}\;\frac{c_{1}^{2j+1}-c_{0}^{2j+1}}{c_{1}-c_{0}},\\
\rho _{j,\alpha } &=& U_{j,\alpha }\rho _{j,1}U_{j,\alpha }^{\dagger } 
 \quad \hbox{for $\alpha\ne 1$},\\
\rho_{j,1} &=&\rho _{j}\otimes (|\Psi _{-}\rangle \langle \Psi _{-}|)^{\otimes J-j},\\
\label{rhoj}
\rho_{j} &=& \frac{c_{1}-c_{0}}{c_{1}^{2j+1}-c_{0}^{2j+1}}
  \sum_{m=-j}^{j}c_{0}^{j-m}c_{1}^{j+m}|j,m\rangle_{\vec n}\langle j,m|.
\end{eqnarray}
\end{mathletters}
This last density operator can be conveniently written as convex sum of
density operators corresponding to identical pure states of $2j$ qubits:
\begin{eqnarray}
\label{rhoj2} 
\rho _{j} &=&\frac{c_{1}-c_{0}}{c_{1}^{2j+1}-c_{0}^{2j+1}}(2j+1)\\
&&\times \int \frac{d\Omega }{4\pi }n(\theta )^{2j}\left( |\Psi (\theta
,\phi )\rangle \langle \Psi (\theta ,\phi |\right)^{\otimes 2j},  \nonumber
\end{eqnarray}
where $n(\theta )=c_{1}\cos ^{2}(\theta /2)+c_{0}\sin ^{2}(\theta /2)$ and 
\begin{equation}
|\Psi (\theta ,\phi )\rangle =
 \sqrt{c_{1}}\frac{\cos (\theta /2)}{\sqrt{n(\theta )}}
 |1\rangle _{\vec{n}}+\sqrt{c_{0}}
 \frac{\sin (\theta /2)}{\sqrt{n(\theta )}}e^{i\phi }|0\rangle _{\vec{n}}
\end{equation}
Any state of the form $\rho^{\otimes N}$ can be decomposed into the
$\rho_{j,\alpha }$ components as in Eq.\ (\ref{rhoN}); this can be seen
by writing $\rho ^{\otimes N}=\sum_{k=0}^{N} c_{0}^{k} c_{1}^{N-k}
\hat{P}_{k}$, where $\hat{P}_{k}$ is a projector on the subspace with
exactly $k$ qubits in state $|0\rangle_{\vec{n}}$. Expressing this
projectors in terms of the basis $|j,m,\alpha\rangle_{\vec n}$
immediately gives (\ref{rhoN}).

Let us now use decomposition (\ref{rhoN}) to introduce our purification
procedure. Given the state $\rho^{\otimes N}$ we perform a measurement
defined by the projections on the mutually orthogonal subspaces
$S_{j,\alpha}$. This provides us with one of the states
$\rho_{j,\alpha}$. If $\alpha \neq 1$ we apply the appropriate unitary
transformation $U_{j,\alpha }^{\dagger }\rho _{j,\alpha }U_{j,\alpha }$
to obtain $\rho_{j,1}$. The resulting density operator, $\rho_{j,1}=
\rho_{j}\otimes (|\Psi _{-}\rangle \langle \Psi _{-}|)^{\otimes J-j}$
gives a clear separation between the first $2j$ qubits in state
$\rho_{j}$ and the remaining qubits paired into $J-j$ singlets. Since
the singlet state carries no information about state $\rho$, we can
discard the last $2(J-j)$ qubits. This procedure gives, with probability
$p_{j}$, $2j$ qubits in state $\rho_{j}$. The reduced density operators
of the resulting qubits coincide and have a fidelity
\begin{equation}
\label{Fj}
f_{j} = \frac{1}{2j}\left[ \frac{(2j+1)c_{1}^{2j+1}}
 {c_{1}^{2j+1}-c_{0}^{2j+1}}-\frac{c_{1}}{c_{1}-c_{0}}\right].  
\end{equation}
Thus, we have that the purification procedure gives [cf. Eq.\ (\ref{PMFM})]
$P_M=p_{j=M/2}$ and $F_M=f_{j=M/2}$ if $M$ is even and $P_M=0$ otherwise.

In order to illustrate our considerations with an example, let us take the
simplest case of $N=2$ qubits. The density operator $\rho ^{\otimes 2}$ can
be decomposed into two terms supported respectively on the singlet
(antisymmetric, $S_{0,1}$) and the triplet (symmetric, $S_{1,1}$)
subspace, 
\begin{eqnarray}
\rho ^{\otimes 2} &=&\sum_{j=0}^{1}p_{j}\rho _{j,1}=
 c_{0}c_{1}|\Psi_{-}\rangle \langle \Psi _{-}|\ + \\
 &&+\left(c_{0}^{2}|0\rangle \langle 0|+c_{0}c_{1}|\Psi _{+}\rangle \langle
 \Psi_{+}|+c_{1}^{2}|1\rangle \langle 1|\right). \nonumber
\end{eqnarray}
Our purification procedure consists of projecting onto the symmetric or
antisymmetric subspaces. After successful projection on the symmetric
subspace, we keep both qubits; after successful projection onto the
antisymmetric subspace, we discard both qubits since the singlet state
does not depend on $\vec{n}$, and therefore the qubits do not carry any
information about the original state. This procedure is symmetric and
gives $P_2(\lambda)=1-c_0c_1=(3+\lambda^2)/4=1-P_0(\lambda)$ and
$F_2(\lambda)=c_1 (1-c_0/2)/(1-c_0c_1)>c_1$. Thus, with probability
$P_2(\lambda)$ we obtain a higher fidelity than the original one.

In general, the average distillability factor (or yield) and the mean
fidelity of our procedure are given respectively by 
\begin{mathletters}
\begin{eqnarray}
D_{N}(\lambda ) &\equiv &\sum_{j=0}^{J}p_{j}\frac{2j}{2J}\simeq \lambda +
\frac{1}{N}\frac{1-\lambda }{\lambda }+O(1/N^{2}), \\
\bar{F}_{N}(\lambda ) &\equiv &\sum_{j=0}^{J}p_{j}f_{j}\simeq 1-\frac{1}{2N}
\frac{1-\lambda }{\lambda ^{2}}+O(1/N^{2}).
\end{eqnarray}
\end{mathletters}
In the limit $N\gg 1$, the mean fidelity tends to one whereas the yield
tends to $\lambda$, the length of the Bloch vector. In fact, one can
check that in this limit the distribution $p_{j}$ becomes narrower and
narrower (that is, the width divided by the mean value tends to zero).
This means that if $N$ is sufficiently large we will basically obtain
$N\lambda$ qubits in the unknown pure state $|1\rangle_{\vec{n}}$. Thus,
in the limit $N\to\infty$ our purification procedure gives the maximum
yield. The question is: how good is this purification for finite $N$? In
fact it is the optimal one!

In order to show that we have an optimal procedure we will first show
that given any other purification procedure ${\cal P}'$ it is always
possible to obtain the same values $P_M'$ and $F_M'$ once we have
applied our procedure ${\cal P}$. Suppose we have applied ${\cal P}$ to
the $N$ qubits, obtained the outcome corresponding to a particular
subspace $S_{j,\alpha}$, applied the corresponding operator
$U_{j,\alpha}^\dagger$, and discarded the last $2(J-j)$ qubits. In order
to obtain the results of the procedure ${\cal P}'$ we just have to add
$J-j$ pairs of qubits in singlet states, apply the operation
$U_{j,\alpha}$ to the whole set, and then apply the procedure ${\cal
P}'$. It is straightforward to show that this whole procedure gives the
same values of $P_M'$ and $F_M'$. The intuitive idea behind all this is
very simple: according to (\ref{rhoN}) the density operator
$\rho^{\otimes N}$ can always be regarded as describing the situation in
which we have prepared with probability $p_j/d_j$ the state
$\rho_{j,\alpha}$. Thus, our measurement only identifies the state which
has been prepared without disturbing it, whereas the operation
$U_{j,\alpha}$ is unitary. Therefore, this procedure is to all practical
purposes reversible.

Now, we complete the proof of optimality by showing that after having
applied ${\cal P}$, it is impossible to increase the averaged fidelity
of the outcoming qubits. The idea is to show that any purification
procedure which produces $1$ qubit out of $2j$ qubits in state
$\rho_{j}$ has at most a fidelity $F_1(\lambda)=f_{j}$ given by
(\ref{Fj}). This automatically implies that there exists no procedure
that produces $M$ qubits (for any $M$) with fidelity higher than $f_{j}$
for if it existed, we could construct a procedure that gives $1$ qubit
of fidelity larger than $f_{j}$ by simply discarding the remaining $M-1$
qubits, which contradicts our previous statement. Therefore, we restrict
ourselves to symmetric procedures that produce one qubit out of $2j$
qubits in a state $\rho_j$. Let us consider a general linear and
completely positive map ${\cal P}_{1}$. It follows from the covariance
condition (\ref{cov}) that for any single qubit in state $|\Psi \rangle$
we have 
\begin{equation}
{\cal P}_{1}\left[ \left( |\Psi \rangle \langle \Psi |\right) ^{\otimes 2j}
\right] =x|\Psi \rangle \langle \Psi |+y|\Psi ^{\perp }\rangle \langle \Psi
^{\perp }|,  \label{cov2}
\end{equation}
where $x,y\geq 0$, $x+y\le 1$ and $\langle \Psi^{\perp }|\Psi \rangle
=0$ (this can be easily proved by showing that the result must commute
with any rotation along the axis defined by the Bloch vector
corresponding to $|\Psi \rangle $). Using Eq.\ (\ref{rhoj2}) and the
fact that ${\cal P}_1$ is linear we can obtain ${\cal P}_1(\rho_j)$ and
express the fidelity of ${\cal P}_1(\rho_j) /{\rm tr}[{\cal
P}_1(\rho_j)]$ in terms of $x$ and $y$. The maximum value occurs for
$(x,y)=(1,0)$ and gives exactly $f_j$, which completes the proof.

Interestingly enough the methods introduced above can be used to study
the optimal cloning \cite{Gi97,Br98,We98} of mixed states. In this case
the goal is to start with $N$ qubits all in the state $\rho$
(\ref{rhop}), and produce exactly $M\ge N$ qubits each in a state as
close as possible to the state $|1\rangle_{\vec{n}}$ \cite{cl-note}.
More specifically, defining the average fidelity as in (\ref{FM}), we
define the optimal cloning procedure ${\cal C}$ as the one which gives a
maximal value of $F_M$. Again, one can easily show that symmetric
procedures ${\cal C}^{s}$ are optimal by using the same four steps as
before. Moreover, since to all practical purposes the purification
procedure defined above is reversible, we can study independently the
optimal cloning of $2j$ qubits in the state $\rho_j$ and at the end
average with the corresponding probabilities $p_j$. Let us then consider
a symmetrical procedure that produces $M$ qubits out of $2j$ qubits and
that is described by a linear trace preserving positive map ${\cal
C}_{j}$. The corresponding fidelity will be $\langle 1_{\vec{n}}| {\rm
tr}_{M-1}[{\cal C}_j(\rho_j)]|1_{\vec n}\rangle$, where ${\rm tr}_{M-1}$
denotes the trace over any $M-1$ particles. Using the property
(\ref{cov}) we have
\begin{equation}
\label{prop2}
{\rm tr}_{M-1}[{\cal C}_j(|\Psi\rangle\langle\Psi|^{\otimes 2j})]=
x |\Psi\rangle\langle\Psi| + (1-x) |\Psi^\perp\rangle\langle\Psi^\perp|,
\end{equation}
where 
\begin{equation}
\label{Fpure}
x \le F_{2j,M}^{\rm pur} \equiv \frac{M(2j+1)+2j}{M(2j+2)}
\end{equation}
since otherwise ${\cal C}_j$ would give for pure states better clones
than the optimal cloning procedure \cite{Gi97}. Using the decomposition
(\ref{rhoj2}), linearity, and (\ref{prop2}) we obtain that the optimal
value occurs for the largest value of $x$, i.e. for $x=F_{2j,M}^{\rm
pure}$. Thus, the fidelity of the optimal cloning procedure for mixed
states is
\begin{equation}
F_{N,M}^{\rm mix} = \sum_{j=0}^J p_j \left[ F_{2j,M}^{\rm pur} f_j 
 + (1-F_{2j,M}^{\rm pur})(1-f_j)\right],
\end{equation}
where $p_j$, $f_j$, and $F_{2j,M}^{\rm pur}$ are given in (\ref{Pj}),
(\ref{Fj}), and (\ref{Fpure}) respectively. Defining $\lambda_{N,M}^{\rm
mix}= 2F_{N,M}^{\rm mix}-1$ we obtain $\lambda_{N,M}^{\rm
mix}=\lambda_{N,\infty}^{\rm mix} (M+2)/M$, where
\begin{equation}
\lambda_{N,\infty}^{\rm mix} = \sum_{j=0}^{J} p_j (2f_j-1)\frac{j}{j+1} 
\equiv 2F_{N,\infty}^{\rm mix}-1,
\end{equation}
with $F_{N,\infty}^{\rm mix}$ is the maximal fidelity in the $N\to
\infty$ cloning of mixed states. Actually, using the same arguments as
in \cite{Br98} one can show that this fidelity is the maximal fidelity
with which one can estimate the state $|1_{\vec n}\rangle$ starting from
$N$ copies of the state $\rho$. This way of estimating copies of mixed
states might be interesting in the following context: Alice is sending
$N$ qubits in an unknown pure state to Bob via the depolarizing channel
and Bob's goal is to determine the state that Alice is sending. In Fig.\ 1
we have plotted $\lambda_{N,\infty}^{\rm mix}$ as a function of $N$ for
various values of the length of the Bloch vector. 

In conclusion - in this paper we have introduced a new decomposition of
the multiqubit states of the form $\rho ^{\otimes N}$. This
decomposition was instrumental in constructing the optimal single qubit
purification procedure (and proving that it is indeed optimal) and in
extending the remit of acceptable input states for the optimal quantum
cloners. The decomposition, apart from having applications in restoring
quantum superpositions via the single qubit purification, seems to be a
useful mathematical tool on its own.

\begin{figure}[tbp]
\epsfig{file=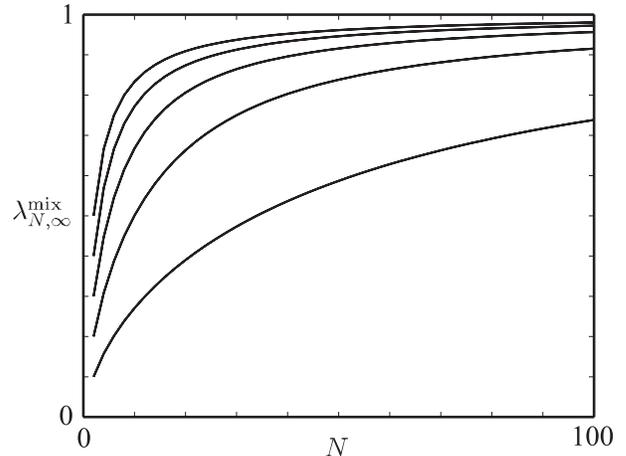,width=8cm}
\caption{Maximum achievable length of the Bloch vector 
$\lambda_{N,\infty}^{\rm mix}$ in the cloning $N\to\infty$ as function of 
the initial number of copies of the state $\rho$. The values of the
initial lengths of the Bloch vector are, from bottom to top, $\lambda=
0.2,0.4,0.6,0.8,1$.
}
\end{figure}

Work supported in part by the \"Osterreichischer Fonds zur FWF,
by the European TMR network ERB-FMRX-CT96-0087, Hewlett-Packard, The
Royal Society London and Elsag--Bailey. After completing this work we
learned that Tarrach {\it et al.} (unpublished) have used a similar
decomposition of the density operator $\rho^{\otimes N}$ to study the
problem of optimal measurements.


\begin{references}
\bibitem{cl-note} 
Note that this scenario is not equivalent to the problem of generating
$M$ output states as close as possible to the mixed state $\rho$.
\end{references}
\end{document}